\documentclass[a4paper,11pt]{article}
\usepackage{pos}

\title{Early time dynamics far from equilibrium via holography}

\author*[a]{Matthias Kaminski}
\author[a,b]{Casey Cartwright}
\author[a]{Marco Knipfer}
\author[c,d]{Michael F.~Wondrak}
\author[e]{Bj\"orn Schenke}
\author[f,g]{Marcus Bleicher}

\affiliation[a]{Department of Physics and Astronomy, University of Alabama, 514 University Boulevard, Tuscaloosa, AL 35487, USA}

\affiliation[b]{Institute for Theoretical Physics, Utrecht University, Princetonplein 5, 3584 CC Utrecht, The Netherlands}

\affiliation[c]{Department of Mathematics/IMAPP, Radboud University, P.O. Box 9010, 6500 GL Nijmegen, The Netherlands}
\affiliation[d]{Department for Astrophysics/IMAPP, Radboud University, P.O. Box 9010, 6500 GL Nijmegen, The Netherlands}

\affiliation[e]{Physics Department, Brookhaven National Laboratory, Upton, NY 11973, USA}

\affiliation[f]{Institut f\"ur Theoretische Physik, Johann Wolfgang Goethe-Universit\"at Frankfurt am Main, Max-von-Laue-Stra{\ss}e 1, 60438 Frankfurt am Main, Germany}
\affiliation[g]{Helmholtz Research Academy Hesse for FAIR (HFHF), GSI Helmholtz Center for Heavy Ion Physics, Campus Frankfurt, Max-von-Laue-Str. 12, 60438 Frankfurt am Main, Germany}

\emailAdd{mski@ua.edu}
\emailAdd{c.c.cartwright@uu.nl}
\emailAdd{mknipfer@crimson.ua.edu}
\emailAdd{m.wondrak@astro.ru.nl}
\emailAdd{bschenke@bnl.gov}
\emailAdd{bleicher@itp.uni-frankfurt.de}

\abstract{
We investigate the early time dynamics of heavy ion collisions studying the time evolution of the energy-momentum tensor as well as energy-momentum correlations within a uniformly thermalizing holographic QGP. From these quantities, we suggest a far-from equilibrium definition of shear viscosity, which is a crucial property of QCD matter as it significantly determines the generation of elliptic flow already at early times. During an exemplary initial heating phase of the holographic QGP the shear viscosity of entropy density ratio decreases down to 60\%, followed by an overshoot to 110\% of the near-equilibrium value, $\eta/s=1/(4\pi)$. 
Implications for the QCD QGP are discussed. 
Subsequently, we consider a holographic QGP which is Bjorken-expanding. Its energy-momentum tensor components have a known hydrodynamic attractor to which all time evolutions collapse independent of the initial conditions. Based on this, we propose a definition for a far from equilibrium speed of sound, and analytically compute its hydrodynamic attractor. 
Subjecting this Bjorken-expanding plasma to an external magnetic field and an axial chemical potential, we study the chiral magnetic effect far from equilibrium. 
}

\FullConference{ HardProbes2023\\
 26-31 March 2023 \\
 Aschaffenburg, Germany\\}

\begin{document}
\maketitle

\section{Introduction}
\label{sec:introduction}
One important practical and theoretical question is why relativistic hydrodynamics describes heavy-ion collision data far beyond its regime of applicability. In particular, hydrodynamics appears to be a valid description far away from local and global equilibrium, in the presence of large gradients, at very early times during the evolution of quark-gluon-plasma (QGP) after collisions of heavy ions or even heavy-light (Pb+p) and light-light (p+p) collisions~\cite{Romatschke:2017ejr}. In part, these points were confirmed in holographic plasma~\cite{Chesler:2010bi} in which numerical computation of all observables is possible at all times. 
Here, we report on the continued holographic exploration of the far-from-equilibrium regime of $\mathcal{N}=4$ Super-Yang-Mills (SYM) theory. We use the holographic correspondence to compute three time-dependent quantities: the shear transport, the speed of sound, and the chiral magnetic current.

\section{$\eta/s$ far from equilibrium}
\label{sec:FFEShear}
We intend to explore the early times after a heavy-ion collision during which the system is far from equilibrium. 
Near equilibrium, a Kubo formula relates the retarded momentum space shear correlator $\tilde{G}_R^{xy,xy} =\langle T^{xy} T^{xy} \rangle$ at vanishing spatial momentum to the shear viscosity: $\eta = -\lim_{\omega\to 0}\frac{1}{\omega} \mathrm{Im} \, \tilde{G}_R^{xy,xy} (\omega,\mathbf{k}=\mathbf{0})$. 
Here, we holographically compute $\tilde{G}_R^{xy,xy}$ far from equilibrium and define a {\it far-from-equililbrium shear viscosity}~\cite{Wondrak:2020tzt}
\begin{equation}
\label{eq:FFEShear} 
\eta(t_{avg}) = -\lim_{\omega\to 0}\frac{1}{\omega} \mathrm{Im} \, \tilde{G}_R^{xy,xy} (t_{avg}, \omega,\mathbf{k}=\mathbf{0})\, ,
\end{equation} 
where $t_{avg}$ is the time with which the state changes as discussed below. 

\begin{figure}[tb]
\center
\includegraphics[width=.49\linewidth]{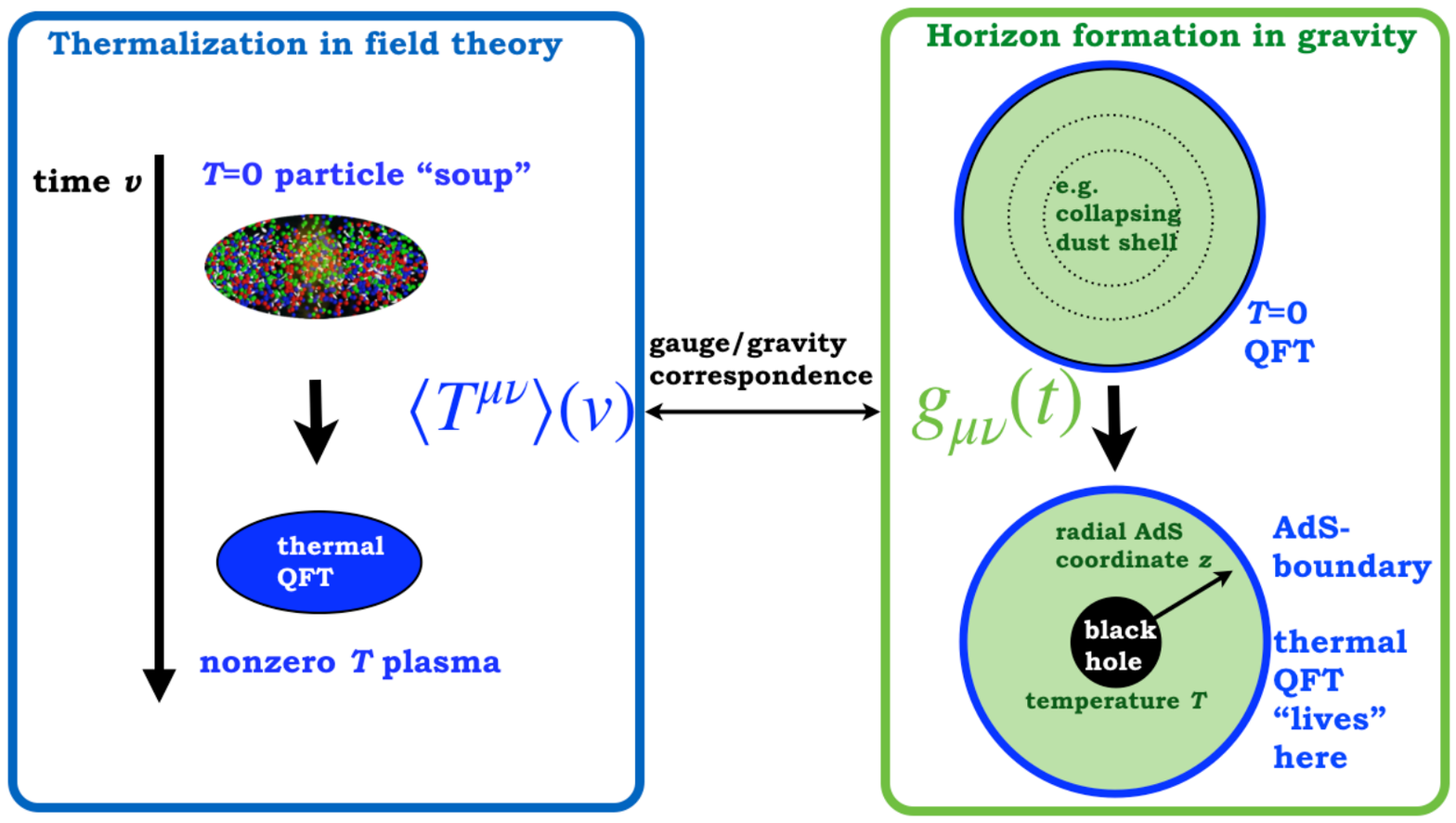}
\hfill
\includegraphics[width=.49\linewidth]{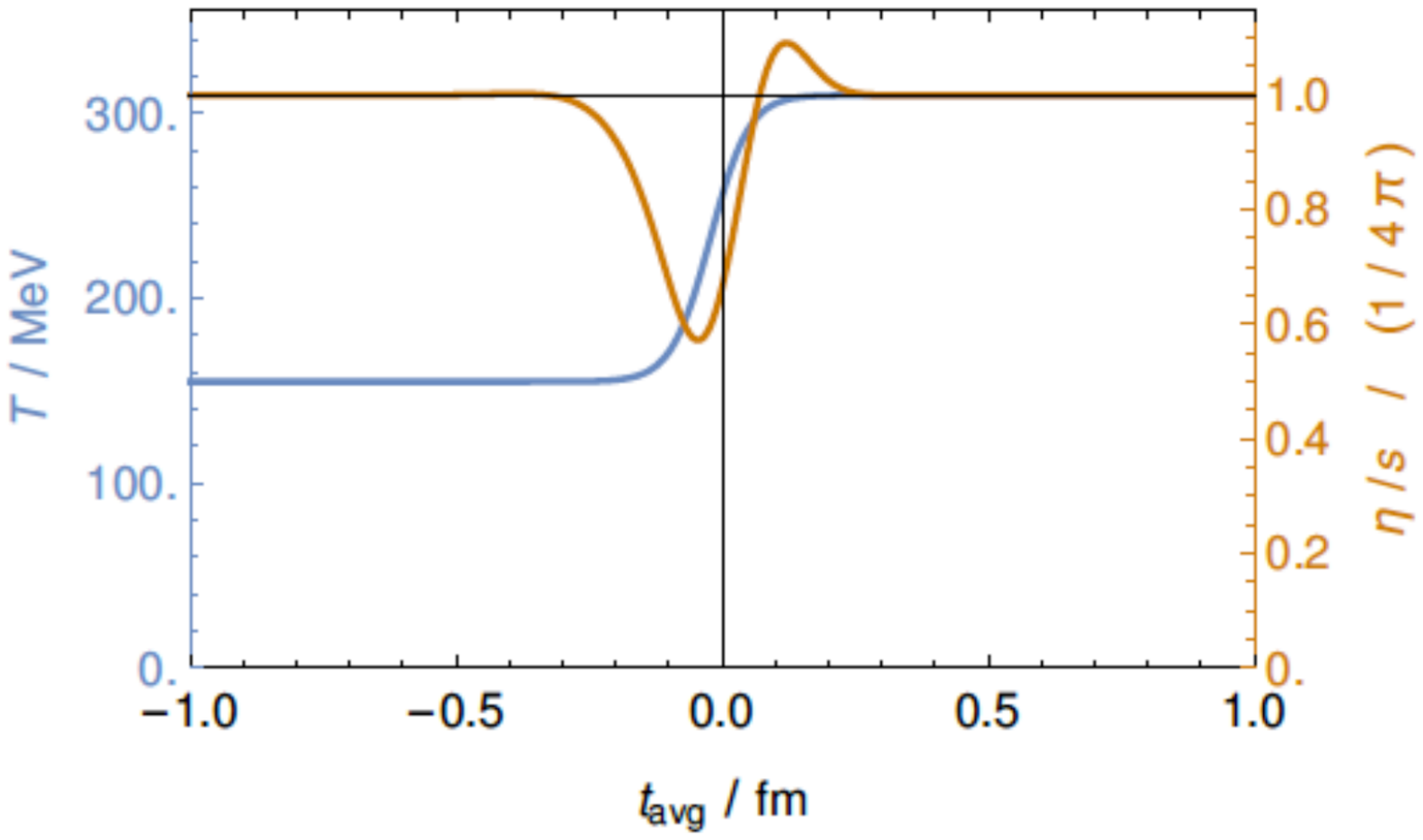}

\caption{\label{fig:FFEShear} 
Left: Thermalization in quantum field theories corresponds to horizon formation in their gravity dual description. 
Right: Example for a time-evolution of the far-from-equilibrium shear~\eqref{eq:FFEShear} normalized to the entropy measure $s(t_{avg})\propto \frac{\partial S^\mathrm{on-shell}}{\partial T}$ based on identification of $S^{\mathrm{on-shell}}$ with the generating functional. 
}
\end{figure}

%
Thermalization of a plasma corresponds to horizon formation in the gravity dual~\cite{Janik:2005zt}, Fig.~\ref{fig:FFEShear} (left). 
A far-from-equilibrium plasma state heating up over a time $\Delta t$ is modeled~\cite{Wondrak:2020tzt} by the AdS$_4$ Vaidya metric
\begin{equation}
\label{eq:VaidyaMetric}
ds^2=g_{\mu\nu}dx^\mu dx^\nu=\frac{1}{z^2} (-f(t,z) dt^2-2 dt dz + dx^2 + dy^2) \, , \quad f(t,z)=1-2G_N M(t) z^3 \, ,
\end{equation}
with the time coordinate $t$, the radial AdS-coordinate $z$, having the boundary at $z=0$ and the horizon at $z=1$, and Newton's gravitational constant $G_N$. Note, that the black hole mass $M(t)=m+m_s(1+\tanh (t/\Delta t))/2$ is a function of the time $t$.\footnote{This metric is a solution of Einstein's equations for any differentiable function $M(t)$. Compared to~\cite{Wondrak:2020tzt} (bulk time $v$), here we chose a different naming convention for the time in the five-dimensional gravity spacetime (here, bulk time $t$).}  
The background metric~\eqref{eq:VaidyaMetric} is perturbed by a metric shear perturbation, $h_{xy}(t,z)$, which is required to solve a linearized Einstein equation. Solutions correspond to the expectation value of the energy-momentum tensor of the plasma and its source $h_{\mu\nu}^{(0)}$ according to ${ h_{\mu\nu}}  \sim { h^{(0)}_{\mu\nu}} + {\langle T_{\mu\nu} \rangle} \, z^4 + \dots$. 
In order to obtain the retarded shear correlator $G_R^{xy,xy}$, linear response theory allows to utilize a delta source: $h_{xy}^{(0)}=\delta(\tau-t_p)$. This yields the two-point function in terms of a one-point function at time $t$ in presence of a delta-source at time $t_p$: $\langle T^{xy}\rangle_{\delta(t_p)}=\int d\tau G^{xy,xy}_R (\tau, t) \delta(\tau-t_p)\propto G^{xy,xy}_R (t_p, t)$. Assuming no dependence on spatial boundary coordinates $x$ or $y$, a Wigner transform now yields the representation in terms of the relative frequency $\omega$: $G_\mathrm{R}^{xy,xy} (t_\mathrm{p},t) \to  G_\mathrm{R}^{xy,xy} (t_\mathrm{avg},t_\mathrm{rel}) \sim \tilde{G}_\mathrm{R}^{xy,xy} (t_\mathrm{avg}, \omega) \, e^{-i\omega t_\mathrm{rel}}$, where the average time is $t_{avg}= (t_p+t)/2$ and the relative time is $t_{rel}=t_p-t$. 

\begin{figure}[tb]
\center
\includegraphics[width=.49\linewidth]{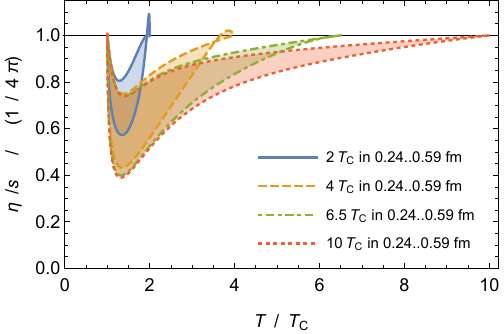}
\hfill
\includegraphics[width=.49\linewidth]{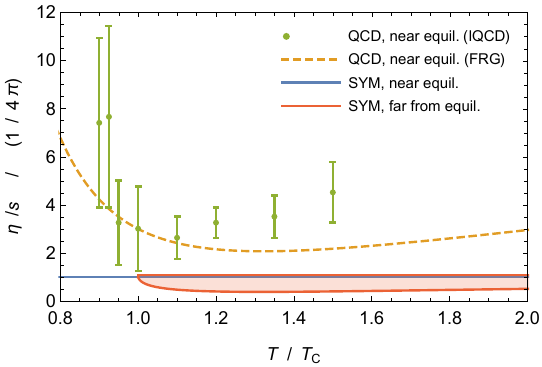}
\caption{\label{fig:etaOfT} 
Left: Dependence of~$\eta/s$ on the instantaneous temperature, $T(t_{avg})$ defined from the Hawking temperature of the black hole at each time. 
Shaded areas indicate the values arising from a sweep over a range of heat-up times for a fixed peak temperature. 
Right: These same holographic results~(SYM, area enclosed by red solid curve) compared to $1/(4\pi)$ (SYM, blue line). Theoretical QCD results are computed near equilibrium by functional renormalization group (FRG, dashed)~\cite{Christiansen:2014ypa} and lattice QCD (lQCD, circles)~\cite{Astrakhantsev:2017nrs}. 
}
\end{figure}

%
Near an equilibrium state, the ratio of shear viscosity to entropy density is $\eta/s=1/(4\pi)$ in $\mathcal{N}=4$ SYM theory~\cite{Kovtun:2004de} (black line in Fig~\ref{fig:FFEShear}, right). 
In Fig.~\ref{fig:FFEShear} (right), the shear viscosity~\eqref{eq:FFEShear} is shown for an example plasma heat up starting at $T_c=155$ MeV, ending at $T_{final}=310$ MeV, rising over $\Delta t=0.3$ fm (RHIC energies).  
For this example, the shear transport ratio first drops below 60\%, then rises above 110\% of $1/(4\pi)$.\footnote{It is important to recall that $\eta/s=1/(4\pi)$ is {\it not} a universal lower bound~\cite{Buchel:2008vz}.} 
How typical is this behavior when changing $\Delta t$ and $T_{final}$? Fig.~\ref{fig:etaOfT} (left) shows that over a wide range of values a significant decrease below $1/(4\pi)$ is generic. The increase above $1/(4\pi)$ only exists for small enough $T_{final}<6.5 T_c$.  
Fig.~\ref{fig:etaOfT} (right) shows a stark contrast between the holographic {\it far-from-equilibrium} results ($\eta/s< 1/(4\pi)$), and the {\it near-equilibrium} lattice QCD and {\it near-equilibrium} FRG results (suggesting $\eta/s> 1/(4\pi)$). This may indicate that the Bayesian study~\cite{Bernhard:2019bmu} underestimated the elliptic flow generated at early times. 
%
%

\section{Speed of sound far from equilibrium}
\begin{figure}[tb]
\center
\includegraphics[width=.5\linewidth]{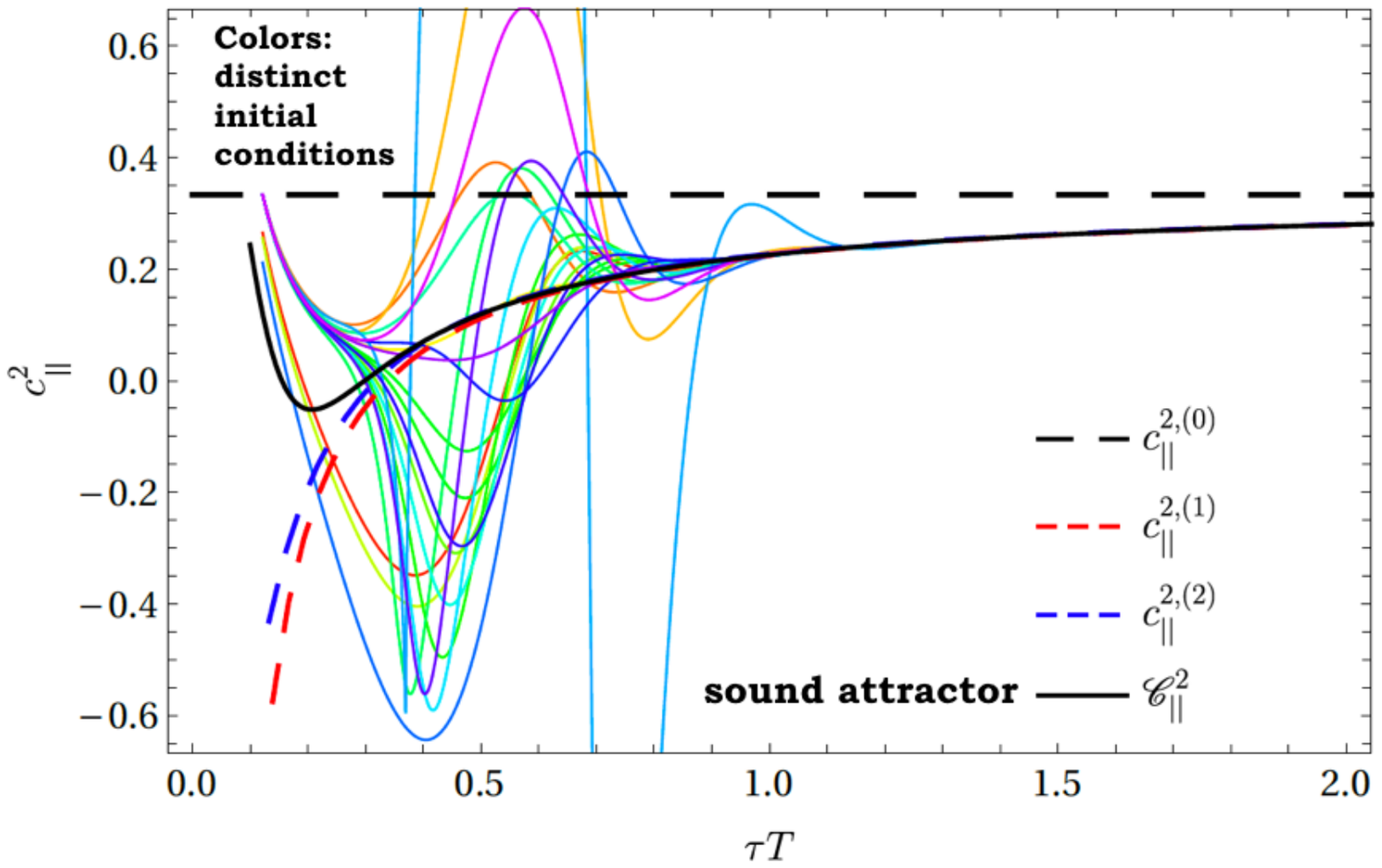}
\caption{\label{fig:soundAttractor} 
Attractor for the longitudinal speed of sound (black solid line) towards which all initial conditions (solid lines, distinct colors) evolve. Dashed: 0th (black), 1st (red), 2nd (blue) order hydrodynamic expectation. 
}
\end{figure}

Consider a Bjorken-expanding $\mathcal{N}=4$ SYM plasma. At early times, thermodynamic quantities are not strictly well-defined as the plasma is far from equilibrium and has a large pressure anisotropy. Here, we propose working definitions far from equilibrium. We use the temperature definition $T = (\epsilon/\sigma_{SB})^{1/4}$, which is sometimes called {\it pseudo temperature}~\cite{Romatschke:2017ejr}. 
We holographically compute the speed of sound far from equilibrium according to the proposed definition~\cite{Cartwright:2022hlg}
\begin{equation}\label{eq:c}
c_{\perp}^2 = - \frac{\partial \langle T^{x_1}_{x_1}\rangle}{\partial \langle T^0_0 \rangle} \, , \quad
c_{||}^2 = - \frac{\partial \langle T^\xi_\xi\rangle}{\partial \langle T^0_0 \rangle} \, ,
\end{equation}
with the pseudorapidity $\xi=\frac{1}{2}\ln [(t+x_3)/(t-x_3)]$, the spatial coordinates $x_1, x_2, x_3$ and the proper time $\tau=\sqrt{t^2-x_3^2}$ with the Stefan-Boltzmann constant $\sigma_{SB}$~\cite{Cartwright:2022hlg}. 
%
Similar to the previous section, a time-dependent metric provides the thermalizing plasma state. However, now this plasma is expanding in the longitudinal $x_3$-direction, while isotropic and uniform in the transverse $(x_1,x_2)$-plane. This complication now only allows numerical solutions for the background metric describing the time-dependent state, using~\cite{Chesler:2010bi}.
%
%
It can be analytically shown~\cite{Cartwright:2022hlg} that the pressure anisotropy attractor~\cite{Spalinski:2017mel} implies an attractor for the time-dependent speed of sound
\begin{equation}
\mathcal{C}^2_{||} = \frac{1}{3} - \frac{2}{9} \left (
\mathcal{A}_0(w)+\frac{w}{4} \frac{\partial \mathcal{A}_0(w)}{\partial w}
\right ) \, ,
\end{equation}
with $w=\tau T$ and the pressure anisotropy attractor $\mathcal{A}_0(w)=(2530 w-276)/(3975 w^2 - 570 w+120)$~\cite{Spalinski:2017mel}. This sound attractor (solid black line) is shown in Fig.~\ref{fig:soundAttractor} along with the numerically computed speed of sound in Bjorken-expanding holographic plasma, starting from various distinct initial conditions (solid colorful lines) and the hydrodynamic expectations (dashed lines). Hydrodynamic expectations are coinciding with the sound attractor already at very early times ($\tau T \approx 0.5$), indicating again a fast hydrodynamization. All initial states evolve towards the sound attractor very quickly, around $\tau T< 1$. 
The perpendicular speed of sound has an analogous attractor~\cite{Cartwright:2022hlg}.  

\section{Chiral magnetic effect far from equilibrium}
\begin{figure}[tb]
\center
\includegraphics[width=.5\linewidth]{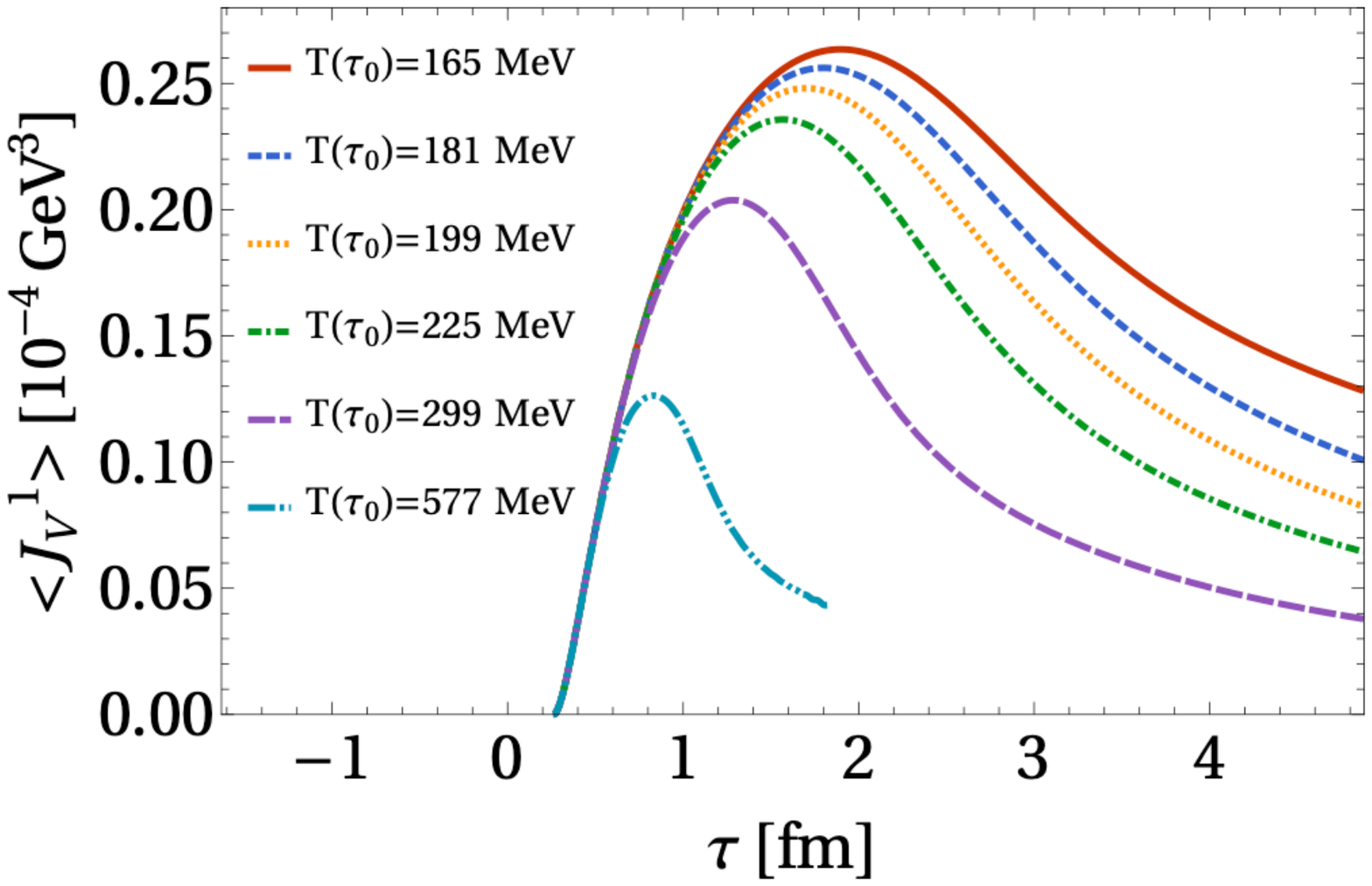}
\caption{\label{fig:CMECurrent} 
The charge current generated along the magnetic field due to the CME at distinct energies ($\propto T^4$). 
}
\end{figure}

In the Bjorken-expanding holographic plasma described in the previous section, we introduce a chemical potential $\mu$ and magnetic field $B$ which both depend on time due to the Bjorken-expansion. In this setting, we compute~\cite{Cartwright:2021maz} (highlighted in~\cite{DOEHighlight2023}) the time-dependent chiral magnetic current $\langle J_V^1\rangle$ generated due to the chiral magnetic effect (CME).
At distinct energies, this current first increases rapidly and then decreases slower, see Fig.~\ref{fig:CMECurrent}. Although Fig.~\ref{fig:CMECurrent} suggest the CME to be weaker at higher energies, the accumulated charge which would be measured in the detectors indicates the opposite to be true when various parameter combinations are considered~\cite{Cartwright:2021maz}. 

\section{Discussion}
We have computed time-dependent shear viscosity, speeds of sound, and the chiral magnetic current in holographic plasmas far from equilibrium. 
A small value of $\eta/s$ at early times implies large generation of elliptic flow at early times, challenging current assumptions. 
In order to check the far-from-equilibrium speed of sound definition~\eqref{eq:c}, the speed of sound waves is to be calculated directly from the
fluctuations around Bjorken-expanding holographic plasma, using techniques from~\cite{Wondrak:2020tzt}. 
For a conclusive CME current estimate, a dynamical magnetic field interacting with the charged plasma, and a dynamically created axial imbalance need to be included. 
In summary, hydrodynamics performs well when its definitions are pushed beyond their limits. This may suggest that an effective field theory of fluid dynamics far from equilibrium is awaiting its construction. \\
%

This work was supported by an Excellence Fellowship from Radboud University (M.F.W.), the U.S.~Department of Energy grant DE-SC0012447 (C.C., M.K., M.K.), and DOE Contract No. DE-SC0012704 (B.P.S.). 

\end{document}